\documentclass[aps,pra,twocolumn,amsmath,amssymb,superscriptaddress]{revtex4-2}

\usepackage{comment}

\usepackage{amsmath,amsfonts,amssymb}
\usepackage{physics}
\usepackage{graphicx}
\usepackage{setspace}
\usepackage{tocloft}
\usepackage[dvipsnames]{xcolor}

\usepackage{graphicx}
\usepackage{subcaption} 
\usepackage{dcolumn}
\usepackage[mathlines]{lineno}
\usepackage{physics}
\usepackage{hyperref}
\usepackage{multirow}
\usepackage{capt-of} 
\usepackage{bm}
\usepackage{epstopdf}
\usepackage{epsfig}
\usepackage{bbold}
\usepackage[normalem]{ulem}
\usepackage{caption}

\usepackage[dvipsnames]{xcolor}

\usepackage{amsmath}

\cftpagenumbersoff{figure}
\cftpagenumbersoff{table} 

\usepackage{CJK}
\usepackage{mathtools}
\usepackage[normalem]{ulem}
\usepackage{dcolumn}
\usepackage[mathlines]{lineno}
\usepackage{hyperref}
\usepackage{bm}
\usepackage{amssymb}
\usepackage{amsmath}
\usepackage{braket}
\usepackage{color}
\usepackage{xcolor}
\usepackage{soul} 
\usepackage{graphics}
\usepackage{bbm}

\begin{document} 
\title{Topological robustness of classical and quantum optical skyrmions in atmospheric turbulence}

\title{Deep diffractive optical neural networks for detecting Skyrmionic topologies of light}

 \author{Hadrian Bezuidenhout}
 \affiliation{School of Physics, University of the Witwatersrand, Private Bag 3, Wits 2050, South Africa}

 \author{Cade Peters}
 \affiliation{School of Physics, University of the Witwatersrand, Private Bag 3, Wits 2050, South Africa}

 \author{Ram Kumar}
 \affiliation{School of Physics, University of the Witwatersrand, Private Bag 3, Wits 2050, South Africa}

\author{Andrew Forbes}
\affiliation{School of Physics, University of the Witwatersrand, Private Bag 3, Wits 2050, South Africa}

\author{Isaac Nape}
\email{isaac.nape@wits.ac.za}
\affiliation{School of Physics, University of the Witwatersrand, Private Bag 3, Wits 2050, South Africa}


\begin{abstract}
\noindent \textbf{Optical Skyrmions are topological forms of structured light with the potential of an infinite encoding alphabet that is immune to disturbance.  This attractive prospect is hindered by the lack of any topological detector, a challenging problem due to the non-orthogonal nature of the topological invariant ($N$). Here we demonstrate the first deterministic detector for Skyrmionic topologies of light using a deep diffractive optical neural network. Our network uses two independent processing channels of 5 diffractive layers each to map incoming topologies to spatially separated Gaussian channels from which $N$ can be detected. We overcome the complexity of the training by using a spatial mode basis rather than pixels, reducing the training variables by $\times 1000$ compared to current methods.  We use the detector on an input set of 81 input topologies, showing high accuracy even in the presence of significant levels of noise. Finally, to show the practical utility of the device, we transmit and receive an image encoded in a 14-level topological alphabet with no discernible cross-talk. Our work offers a new paradigm for the emergent field of diffractive optical networks and can easily be extended to other forms of optical topologies, setting a clear pathway for their deployment in real-world applications. }
\end{abstract}
 
\maketitle

\noindent The ability to shape light's many degrees of freedom (DoFs) has allowed for the creation of a myriad of complex and exotic structures \cite{forbes2021structured}. Notable among these are optical skyrmions, particle-like topologies which have seen a surge of interest in recent years \cite{shen2024optical}, driven by their now proven robustness to external perturbations in both the classical \cite{wang2024topological,peters2025seeing,guo2025topological} and quantum regimes \cite{ornelas2024non,ornelas2025topological}. The potential applications of these states of light are vast, ranging from communication \cite{peters2025seeing} and information processing \cite{wang2025perturbation} to metrology \cite{dai2020plasmonic} and microscopy \cite{du2019deep}. This enthusiasm has led to the development of a multitude of means to generate these topologies, including static and tunable liquid crystal optics and wave plates \cite{hakobyan2024unitary,hakobyan2025q,koni2025dual}, microcavities \cite{lin2021microcavity}, interferometers combined with digital holograms \cite{gutierrez2021optical,sugic2021particle,shen2022generation,singh2023synthetic,teng2023physical} and advanced metasurface devices \cite{li2024realization,he2024optical}.\\

Despite this large and rapidly growing body of work, the toolkit for manipulating and detecting optical skyrmions is still in its infancy, relying on reconstructions of the underlying optical fields through Stokes polarimetry \cite{singh2020digital} or quantum state tomography \cite{toninelli2019concepts}, both requiring multiple measurements that are highly susceptible to noise. Advancements in this area have been held back by the fact that skyrmions of different topologies are not necessarily orthogonal. This is in stark contrast to light's many other DoFs, such as  polarisation, orbital angular momentum (OAM) and wavelength to name a few, which do form orthogonal bases and whose elements are easy to discriminate between with the use of single outcome probabilistic  \cite{roux2014projective} or multi-outcome deterministic projective measurements \cite{ndagano2017deterministic, mirhosseini2013efficient, defienne2020arbitrary}.\\

\begin{figure*}[t]
	\centering
\includegraphics[width=\linewidth]{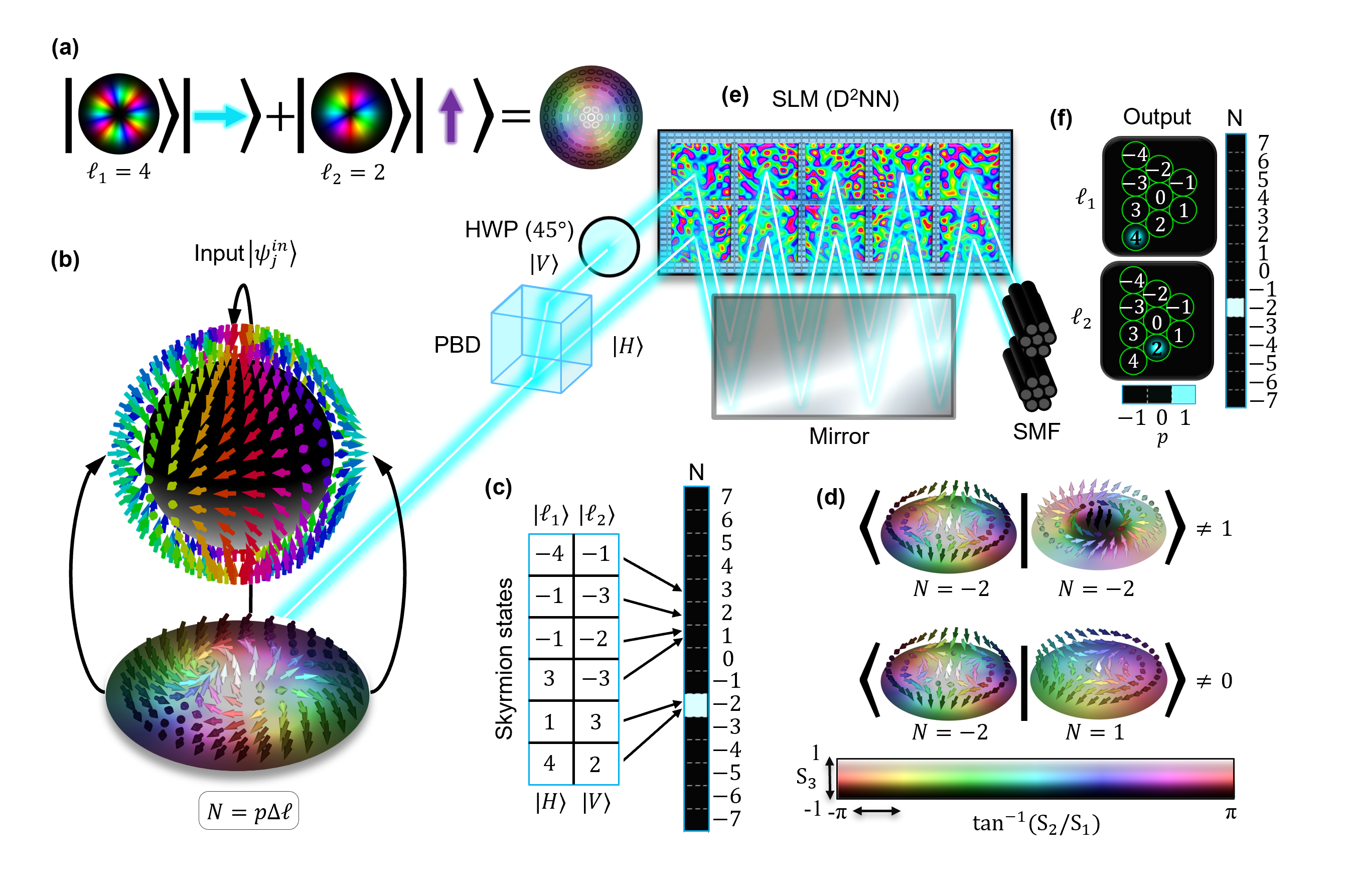}
	\caption{\textbf{Concept of a deterministic detector for optical Skyrmions.}
(a) Vector beam carrying skyrmion topology with orbital-angular-momentum (OAM) of $\ell_{1,2}$ and polarisation states, $H/V$. 
(b) Each vector beam defines a spatially varying Stokes spin field $\mathbf{S}(\mathbf{r})$ in the transverse plane, 
which is mapped onto the unit sphere via inverse stereographic projection, establishing the topological mapping 
$\mathcal{R}^2 \rightarrow \mathcal{S}^2$. The Skyrmion number $N =p \Delta\ell$ depends only on the polarity 
$p=\text{sgn}(|\ell_1|-|\ell_2|)$ and the vorticity $\Delta\ell=\ell_2-\ell_1$.
(c) Multiple vector beams can therefore belong to the same topological class, making the mapping from vector modes to Skyrmion classes \emph{surjective}.
For example, the beams $(\ell_1,\ell_2)=(4,2)$ and $(1,3)$ both yield $N=-2$ despite being orthogonal. 
(d) Conversely, distinct Skyrmion classes can contain overlapping or non-overlapping vector modes. 
(e) A diffractive neural network that resolves the OAM for each polarisation component in a single shot so that we can leverage the OAM/polarisation information to determine (f) the polarity $p$, vorticity $\Delta \ell$, and thus deterministic classification of $N$.}
    \label{fig:concept_topolgy}
\end{figure*}

Artificial neural networks are well suited tools for solving complex, multi-parameter problems and have been realized using light \cite{sui2020review,zhang2025structured}, termed optical neural networks (ONNs) or deep diffractive neural networks (D$^2$NN), motivated by the decreased energy demands of optical systems, where nature itself is able to take on some of the computational burden \cite{wetzstein2020inference,fu2024optical}. These systems have demonstrated the ability to tackle an array of tasks including image classification \cite{li2021all}, spatial mode (de)multiplexing \cite{huang2021all,huang2022orbital,zhang2022polarized,mei2023cascaded,fontaine2019laguerre,zhou2024spatiotemporal,li2024redefinable}, sorting of non-orthogonal quantum states \cite{goel2023simultaneously}, phase retrieval \cite{mengu2022all}, emulation of quantum gates \cite{wang2024ultrahigh,wang2024polarization}, photonic computing \cite{qian2020performing,zarei2022realization,spall2020fully,koni2024emulating} and noise mitigation \cite{jia2022compensating}.  

Here, we deploy the power of D$^2$NN to demonstrate the first instance of a detector for optical topologies, where an arbitrary input Skyrmion topological state is mapped to two polarisation-dependent output channels through a series of five D$^2$NN masks, reaching detection visibilities (equivalently fidelities) of $>93\%$. The joint outcome on each channel allows for the direct measurement of the topological invariant, the Skyrmion number $N$, which we verify using 81 input topologies across 15 different Skyrmion numbers. To overcome the complexity of the training we adopt a modal basis for the mask reconstruction, using Fourier modes rather than pixels, for a network that requires only $\approx 10^2$ diffractive units per layer rather than the $\approx 10^5$ for pixel-based approaches. To demonstrate the effectiveness of such a detector, we perform a proof-of-principle experiment and use the detector to receive information encoded into a topological alphabet, showing excellent fidelity even under severe channel distortions and high background light levels. Our demonstration of an easy to train, high speed and noise resilient topological detector paves the way for the use of topological light in a wide range of real-world applications, including metrology and sensing, imaging and high-speed classical and quantum optical communications.



\section*{Results}

Our approach for detecting optical Skyrmions using diffractive neural networks is illustrated in Fig.~\ref{fig:concept_topolgy}. The overarching goal is to detect Skyrmions in an all-optical processing scheme while addressing the key challenge that topological numbers are not orthogonal.

\subsection{Concept of topological classifier}

Skyrmions are field configurations and, as illustrated in Fig.~\ref{fig:concept_topolgy} (a), can be realised in paraxial optical fields using nonseparable vector beams of the form \cite{gao2020paraxial},
 \begin{equation}
     \ket{\psi_{\ell_{12}}} = \ket{\ell_1}\ket{H} + \exp(i \gamma)\ket{\ell_2}\ket{V}.
     \label{eq:VM}
 \end{equation}
Here, the first ket $(\ket{\ell_{1(2)}})$ denotes the OAM degree of freedom while the second ket $(\ket{H(V)})$ denotes the polarisation degree of freedom of each photon in the field, while $\gamma$ is a relative phase. In this work, we define the OAM eigenmodes using cylindrically symmetric Laguerre-Gaussian modes which can be expressed in polar coordinates $\bar{r} = (r, \phi)$,  $ \langle \bar{r} | \ell \rangle = \text{LG}_{\ell}\left( \bar{r} \right) \propto \sqrt{2}(r/w)^{|\ell|} \exp(-(r/w)^2) \exp(i \ell \phi)$ where 
$\ell$ is a topological index that is associated with an OAM of $\ell\hbar$ per photon. To reveal the topology, it is essential to represent the vector mode as a spin field, analogous to a space-dependent magnetic spin configuration, using the normalised Stokes vector, $\bar{S}(\bar{r})$ \cite{singh2023synthetic}. This representation maps the spin vectors across a plane, $\mathcal{R}_2$, in real space as illustrated in Fig.~\ref{fig:concept_topolgy} (b). An inverse stereographic projection of the spin textured field maps the field onto a unit sphere so that the polar coordinates ($r, \phi) \in \mathcal{R}^2$ are replaced with the spherical coordinates $(\Phi,\Theta) = (2 \arctan (1/r),\phi) \in \mathcal{S}^2$, mapping the spin field onto the surface of a unit sphere. This represents a mapping from $\mathcal{R}^2 \rightarrow \mathcal{S}^2$. It can be shown that the Stokes spin vector $\bar{S}(\bar{r})$ also serves as a mapping function from $\mathcal{S}^2 \rightarrow \mathcal{S}^2$, from the plane represented as a sphere ($\mathcal{S}^2$) to the parameter space for polarisation states (Poincaré sphere, $\mathcal{S}^2$). Conveniently, one can extract the topological invariant, i.e. the Skyrmion number, quantifying the number of times the spin vectors wrap the Poincaré sphere when tracing out a path in physical space, given by
\begin{equation}
N = \frac{1}{4 \pi} \int_{A} \mathbf{S} \cdot 
\left( 
\frac{\partial \mathbf{S}}{\partial x} \times 
\frac{\partial \mathbf{S}}{\partial y}
\right)\, dx\, dy
\label{eq:SkyrmeNo}
\end{equation}
where $(x, y)$ are the Cartesian coordinates. Given a vector mode embedded with the Skyrmionic topology, as in Eq.~(\ref{eq:VM}), the Skyrmion number can be expressed compactly as
\begin{equation}
N =  p \Delta \ell,
\label{eq:SkyrmeNo2}
\end{equation}
where $p =\text{sgn}\left(|\ell_{1}| - |\ell_{2}|\right)$ defines the polarity, while $\Delta \ell = \ell_2 - \ell_1$ defines the vorticity of the spin field. We now see that the skyrmion number is parametrised by the OAM charge $\ell_{1(2)}$ marking each polarisation component. Furthermore, we can form equivalence classes, $[N_{\ell_{12}}]$, or equivalently, Homotopy groups $N_{\ell_{12}} \in \pi_2(S^2)$, that form collections of all Skyrmions with the same topological number, i.e. that can be smoothly deformed into each other. This means that the mapping from the vectorbeams $(\ket{\psi_{\ell_{12}}})$ to Skyrmion classes is surjective (many-to-one) as shown in Fig.~\ref{fig:concept_topolgy} (c). For example, the Skyrmions corresponding to vector beams with OAM values ($\ell_1 =4, \ell_2=2$) and ($\ell_1 =1, \ell_2=3$) both map onto the same Skyrmions number ($N=-2$) even though they are orthogonal, and so it is not sufficient to construct a measurement device that can distinguish between orthogonal vector modes. Figure~\ref{fig:concept_topolgy} (d) shows, conversely, that Skyrmion classes are not characterised by the inner products between the fields: two optical fields belonging to distinct Skyrmion classes may still have nonzero or even zero overlap and therefore, orthogonality of vector modes does not correspond to their topological distinction. Therefore, because traditional sorting procedures for optical modes make use of orthogonality between states of light through linear/unitary optical elements, it is difficult to detect Skyrmions from the field functions alone since they do not map one-to-one with the Skyrmion number. \\

Nonetheless, we show that one can intelligently leverage the unitarity of mode sorting transformations and information embedded in the polarisation components to solve this problem.  To achieve this, we design a detector that allows us to extract the polarity ($p$) and the vorticity ($\Delta\ell$) through a single-shot measurement. The configuration for achieving this is shown in Fig.~\ref{fig:concept_topolgy} (e). This is done by 1) separating the polarisation components using a beam displacer and 2) resolving the OAM modes that mark them using a diffractive neural network, constructed from cascaded phase masks that can be trained to sort the OAM modes. The resulting beam is then mapped deterministically to a unique position in space, correlating it with the underlying OAM content. Once $\ell_1$ and $\ell_2$ are resolved, the Skyrmion class is then determined as illustrated in Fig.~\ref{fig:concept_topolgy} (f).

\subsection{Diffractive network architecture}
\begin{figure*}[t]
	\centering
\includegraphics[width=\linewidth]{ 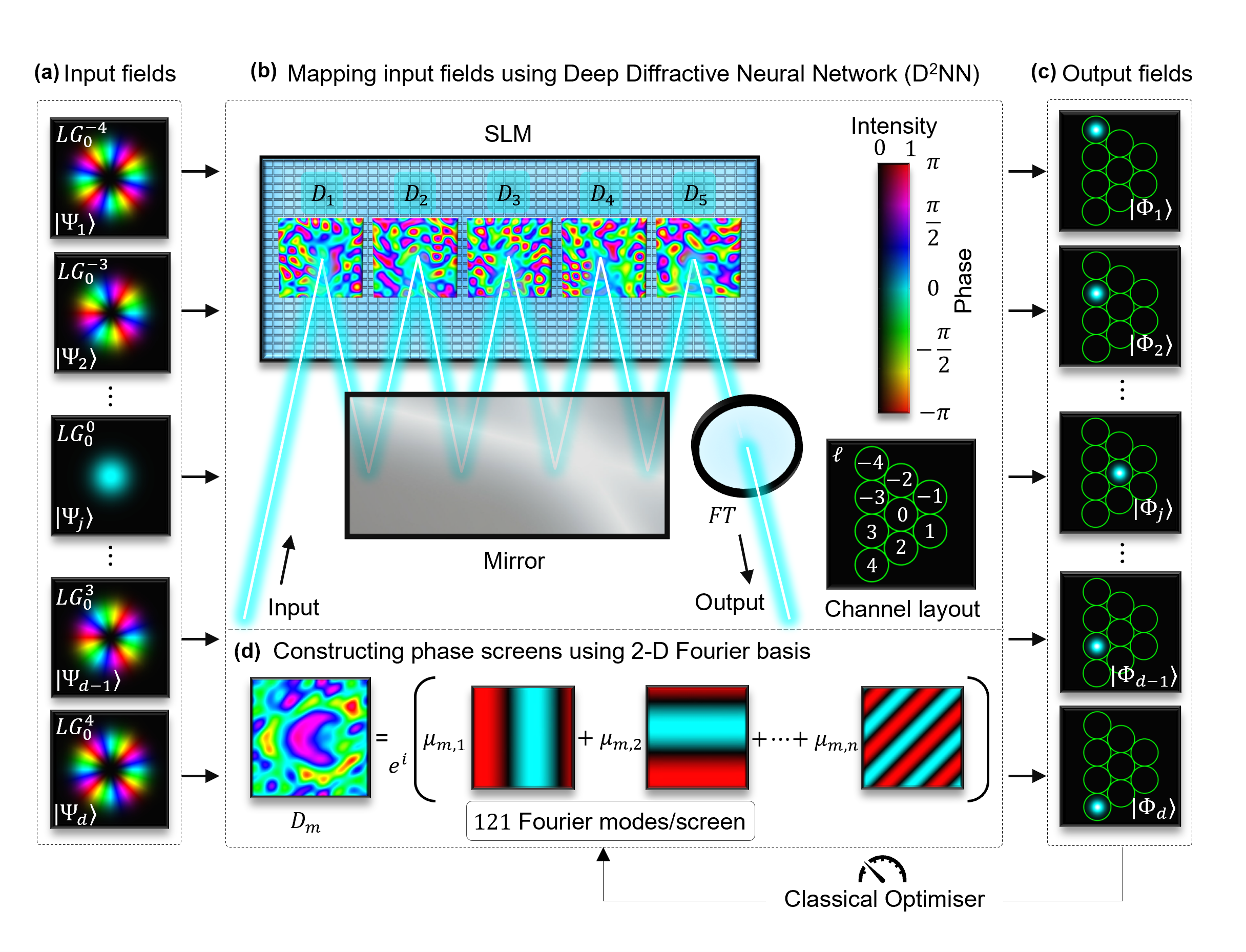}
	\caption{ \textbf{Sorting modes with modes.}
(a) A set of input OAM eigenmodes $\ket{\Psi_j}\equiv\ket{\ell_j}$ is injected into the system, with the goal of mapping each OAM value to a unique spatial location. 
(b) The sorting operation is implemented using a diffractive neural network that realises an approximate unitary transformation $\hat{T}$ composed of cascaded free-space propagation ($P$) and phase-modulation layers ($D_m$). 
(c) The target transformation maps each input OAM mode to a distinct displaced Gaussian output mode $\ket{\Phi_j}$, forming spatially separated detection bins. 
(d) Instead of pixel-wise phase modulation, each diffractive layer is constructed from a weighted superposition of Fourier modes.
The trainable weights $\mu_{m,n}$ are optimised using simultaneous perturbation stochastic approximation (SPSA) to minimise a cost function that penalises incorrect mode-to-bin mapping. 
This architecture enables efficient, all-optical sorting of OAM modes for both polarisation components in the subsequent Skyrmion detection scheme.}
	\label{fig:concept_sorter}
\end{figure*}

\begin{figure*}[t]
	\centering
\includegraphics[width=\linewidth]{ 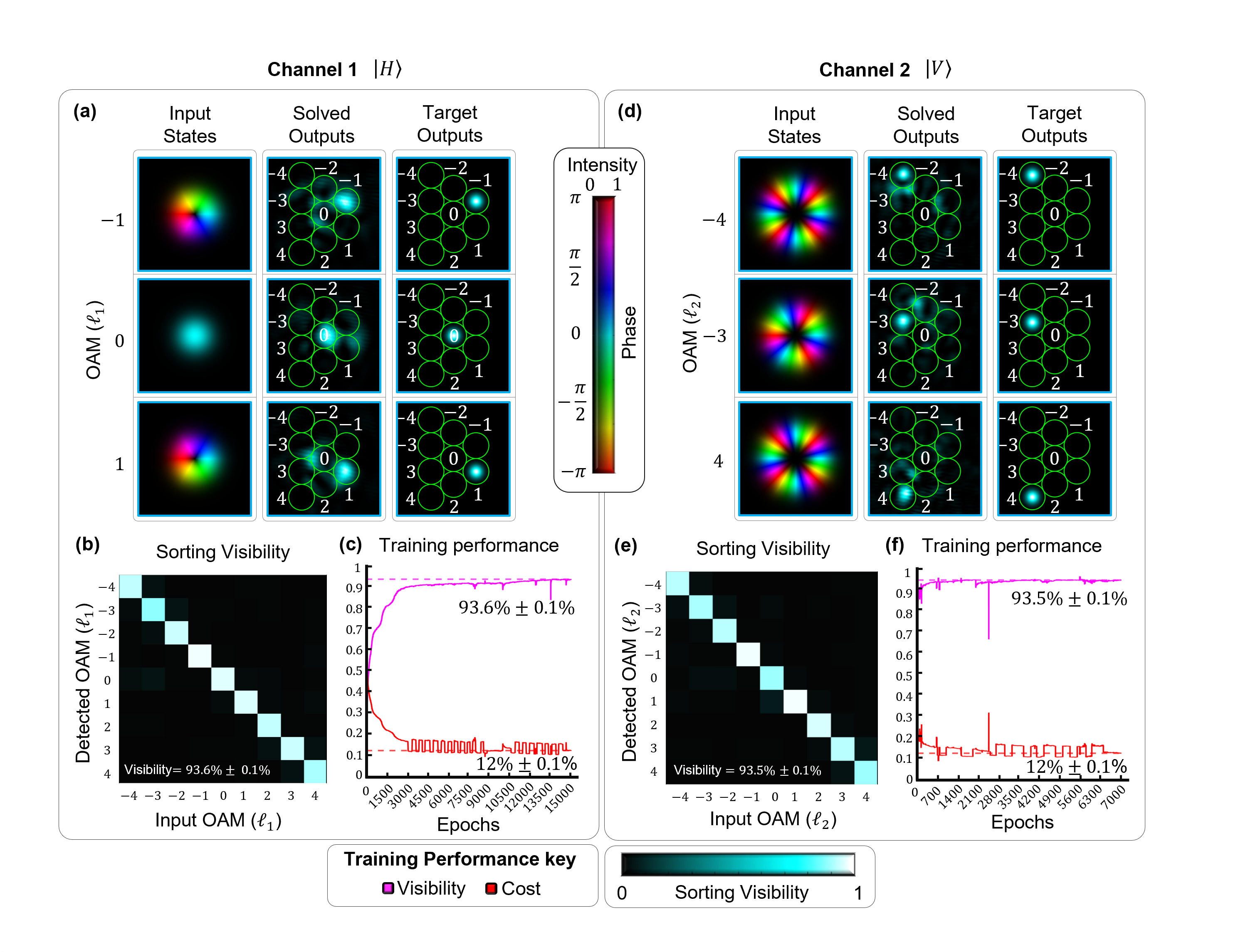}
	\caption{\textbf{Training and validation of the D$2$NN using OAM.} (a) Examples of the simulated input OAM modes and their corresponding target mode is shown alongside the experimentally measured intensity obtained by mapping the input using channel 1 of our D$^2$NN, which will correspond to the horizontally polarised component of our input Skyrmion field. (b) A crosstalk matrix constructed from the normalised inner products between each input mode and all target modes, quantifying the detection probability for each input. (c) The behaviour of the visibility and cost function at each epoch during the phase screen training process, displaying the convergence of the system to its optimum solution. (d) Similar measurement examples are shown for channel 2, corresponding to the vertically polarised component of our input Skyrmion field. (e) Using these projective measurements a crosstalk matrix is again compiled from the normalised visibilities to visualise the probability distribution of the detected modes. (f) The behaviour of the various performance variables at each epoch is again shown for channel 2, where the solved coefficients from channel 1 served as the initial ansatz for the training procedure.}
	\label{fig:results_sorter}
\end{figure*}

To illustrate the working principle of our diffractive network that sorts the OAM states in each polarisation state, as shown in Fig. \ref{fig:concept_sorter}, we first consider a collection of input OAM eigenmodes $\ket{\Psi_j} \equiv | \ell_j \rangle$ as illustrated in Fig. \ref{fig:concept_sorter}. We construct an optical operator $\hat{T}$ using the diffractive network (Fig. \ref{fig:concept_sorter} (b)) mapping $\ket{\Psi_j} \rightarrow \ket{\Phi_j}$, from distinct OAM modes to disjoint locations (bins) of displaced Gaussian modes $\ket{\Phi_j}$ (see Fig. \ref{fig:concept_sorter} (c)). The diffractive neural network follows the universal decomposition for unitary operators
\begin{equation}
\hat{T} \approx P  D_{L-1} .. P D_1 P D_0 P    , \label{eqn:DNN_unitary_approximation}
\end{equation}
having interlaced free-space propagation ($P$) and phase modulation layers ($D_m$) that can be diagonal matrices in the pixel basis. In our approach, we define the phase for each layer using Fourier basis modes $K_{m,n}$, as illustrated in Fig. \ref{fig:concept_sorter} (d), which are a natural periodic basis for describing the constituent frequency components of optical fields and are commonly used to synthesise complex phase mappings \cite{peters2025structured}; forming a complete set of basis functions within a finite real space, spanning the full resolution of each phase screen. Doing this reduces the number of training parameters by a factor of $\sim 10^2$ per phase screen, enabling faster and robust training. Accordingly, each layer in the network was expanded as, 
$D_m(\bar{\mu}_m) \equiv e^{ \circ i \text{Re}{ \{ \sum_{n} \mu_{m,n} K_{m,n}}\}}$ where $e^{\circ}(\cdot)$ denotes the element wise exponential, and $\text{Re}$ denotes the real part. Here, each Fourier basis mode $K_{m,n}$ is ascribed a trainable weighting, $\mu_{m,n} \in $  $\bar{\mu}_m$ that scales its contribution to the construction of the m$^{th}$ layer in the network. The training routine is reported in the Methods section and involves minimising the cost function associated with mapping each input mode to the desired location, via the simultaneous perturbation stochastic approximation (SPSA) approach \cite{ferrie2014_self_guided}. \\

\begin{figure*}[t]
	\centering
\includegraphics[width=\linewidth]{ 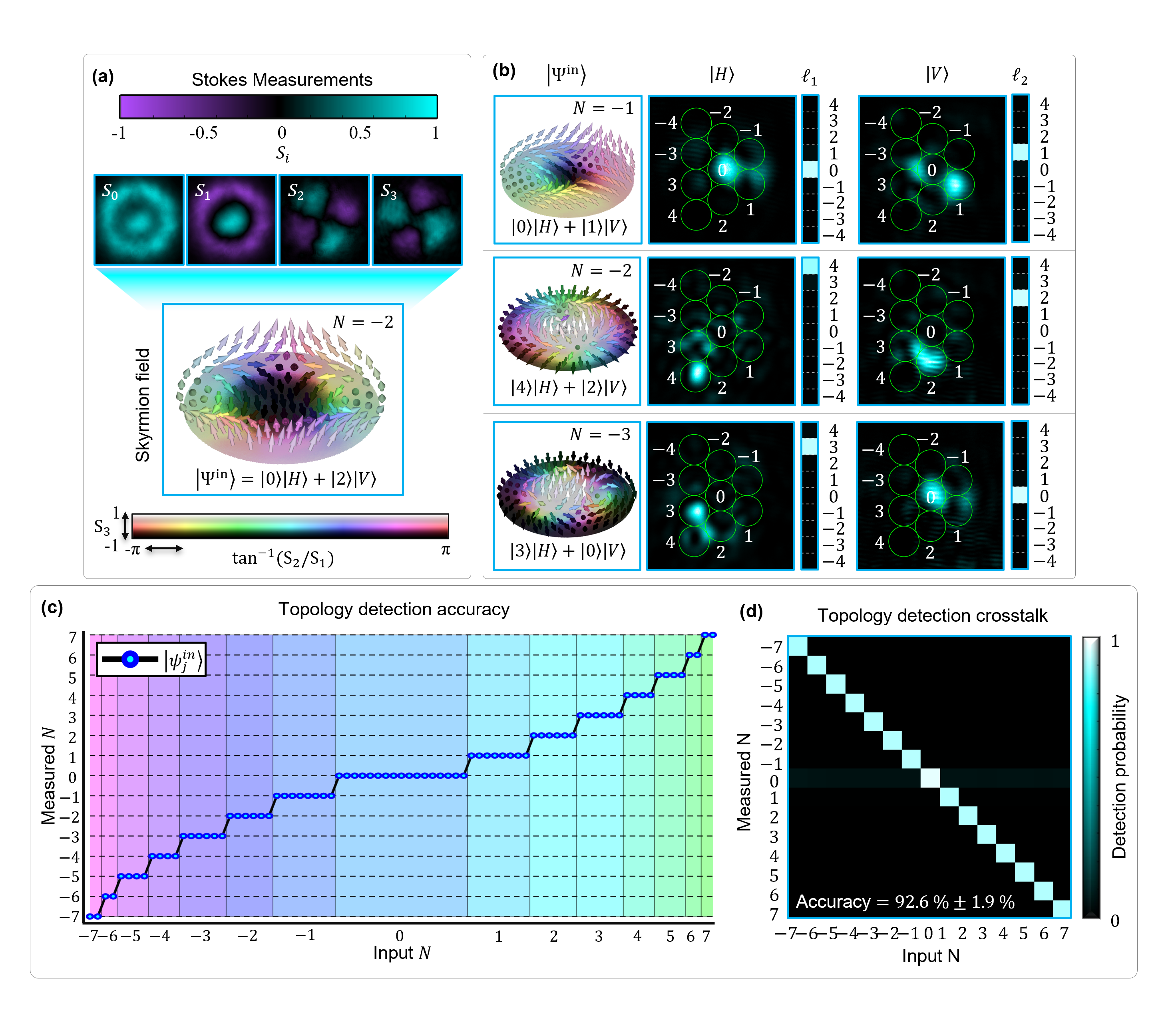}
	\caption{\textbf{Topology detection.} (a) To confirm the quality of the states entering the detector, the polarisation structure of the input Skyrmion fields are reconstructed using a full set of Stokes measurements. (b) The individual $\ket{H}$ and $\ket{V}$ components of the various Skyrmion fields are measured in their respective channel regions, allowing the extraction of their constituent topological charges $\ell_{1,2}$, and the determination of the topological number $N$. (c) Measured topological numbers $N\in[-7,7]$ for $81$ Skyrmion states formed using all possible combinations of $\ell_{1,2}\in[-4,4]$. Multiple input states $\ket{\Psi_j^{\text{in}}}$ within the same class are measured with the same $N$, illustrating a many-to-one (surjective) mapping. (d) A crosstalk matrix representing the detection probability for each topological class, constructed from the measurement of all $81$ Skyrmion states.}
	\label{fig:results_topology}
\end{figure*}

The results of this training procedure are shown in Fig.~\ref{fig:results_sorter}. To construct our topology detector, we were required to train both channel 1 and 2 within our D$^2$NN to classify Laguerre-Gaussian modes based on their topological charges ($\ell_{1(2)}$). The input modes were each assigned a target state defined by spatially separated Gaussian modes that were used to define non-overlapping output channel regions, as shown in Fig.~\ref{fig:results_sorter} (a) and (d) for channels 1 and 2, respectively. The system was trained to accommodate a 9-dimensional mapping within each channel, accounting for values of $\ell_{1(2)}\in[-4,4]$. The experimentally measured intensities were used to construct crosstalk matrices, allowing for the visualisation of the probability distribution of our mode detections, shown in Fig.~\ref{fig:results_sorter} (b) and (e), with each plot representing the normalised visibility of the relative intensity measurements from each channel. The clear diagonal nature of these matrices indicated an accurate set of measurements with little crosstalk, obtaining visibilities of $\approx93\%$ for both channel 1 and 2. These mapping solutions were found using a modified SPSA algorithm, with the behaviour of the visibility, channel efficiency and cost function being shown for each epoch during the training process in Fig.~\ref{fig:results_sorter} (c) and (f). Convergence towards an optimum solution was evident by the steady minimisation of the cost function, accompanied by an increase in the visibility. In both training cases we observed a sharp initial increase in the visibility, with the majority of the training time having been spent increasing the channel efficiency - both channels reached a channel efficiency of $>75\%$. These channels were trained separately by first starting a fresh training run with channel 1 (i.e., $\bar{\mu}=0$) and was allowed to optimise for exactly $15^4$ epochs. The solved coefficients from channel 1 were then used as an initial ansatz for the training of channel 2, allowing for a similar accuracy to be reached with fewer iterations ($7.5\times10^3$ epochs).
\\

To mitigate barren plateaus encountered during optimisation, we incorporated an auxiliary procedure alongside the standard SPSA algorithm. When the convergence rate fell below a defined threshold, the cost landscape was adaptively reshaped by modifying the objective function, akin to curriculum-learning strategies used in machine-learning \cite{Zou2022_Curriculum_Learning,Chng2022_curriculum_keywords,Kim2019__curriculum_learning_strategy,Liu2023_review_curriculum_learning}. This adaptive behaviour manifests as the characteristic ``bumpiness'' observed in the cost-function trajectories of Fig.~\ref{fig:results_sorter}(c) and (f).\\

During training, abrupt spikes and dips in the cost function were also observed, arising from real experimental perturbations such as background-noise fluctuations, transient obstructions in the beam path, or mechanical disturbances to the optical setup. Such events naturally yield erroneous gradient estimates, occasionally pushing the optimisation away from the true optimum. To address this, a fail-safe checkpoint routine was implemented, enabling rapid detection and correction of anomalous updates. The resulting stability is evident by the system’s swift recovery following each disturbance, highlighting the inherent robustness of the approach and underscoring its suitability for training under realistic, imperfect laboratory conditions.

\subsection{Experimental topological classification} 
Having trained both channels to successfully classify OAM modes with high visibility, we moved to classifying full Skyrmion field topologies using single-shot measurements through the D$^2$NN. To confirm that the topology of our input states matched that of the detected states, we measured its polarisation structured using Stokes polarimetry, as shown in Fig. \ref{fig:results_topology} (a). Here the state $\ket{\Psi}=\ket{0}\ket{H}+\ket{2}\ket{V}$ was used as an example, where the Skyrmion number was correctly found to be $N=-2$. Examples of the detector measurements for Skyrmion states of varying topologies are shown in Fig. \ref{fig:results_topology} (b). The two detector regions where the $\ket{H}$ and $\ket{V}$ components were captured were each divided into nine spatially separated channels assigned to the different values of $\ell_{1(2)}$. We constructed a visibility vector from the overlap between each intensity measurement and all target channel states and took the channel with the maximum overlap as the detected value, a method that is commonly used when performing optical classification tasks \cite{li2021all,Ozcan_2018_all_optical_ML}, allowing us to reconstruct the input state and predict its Skyrmion number. \\

As shown in Fig.~\ref{fig:results_sorter}, each channel was trained to map Laguerre-Gaussian modes with topological charges ranging for $\ell_{1(2)}\in[-4,4]$, with the channels corresponding to the $\ket{H}$ and $\ket{V}$ components of our input Skyrmion fields. This OAM range allowed us to construct $81$ unique state combinations in the form of Eq. \ref{eq:VM}, with their Skyrmion numbers spanning $N\in [-7,7]$. All our input states were sent to the detector one after the other and their topologies measured, allowing us to classifying their Skyrmion number as shown in Fig. \ref{fig:results_topology} (c). Multiple unique state combinations being mapped correctly into their relevant topological classes showcased that our detector did not discriminate between states with different geometries within the same class. This is further illustrated by the crosstalk matrix in Fig. \ref{fig:results_topology} (d), indicating that the input states were detected with very high fidelity.
\subsection*{Image reconstruction}
\label{sec:image_reconstruction}
\begin{figure} [h!]
   \centering
   \includegraphics[width=\linewidth]{ 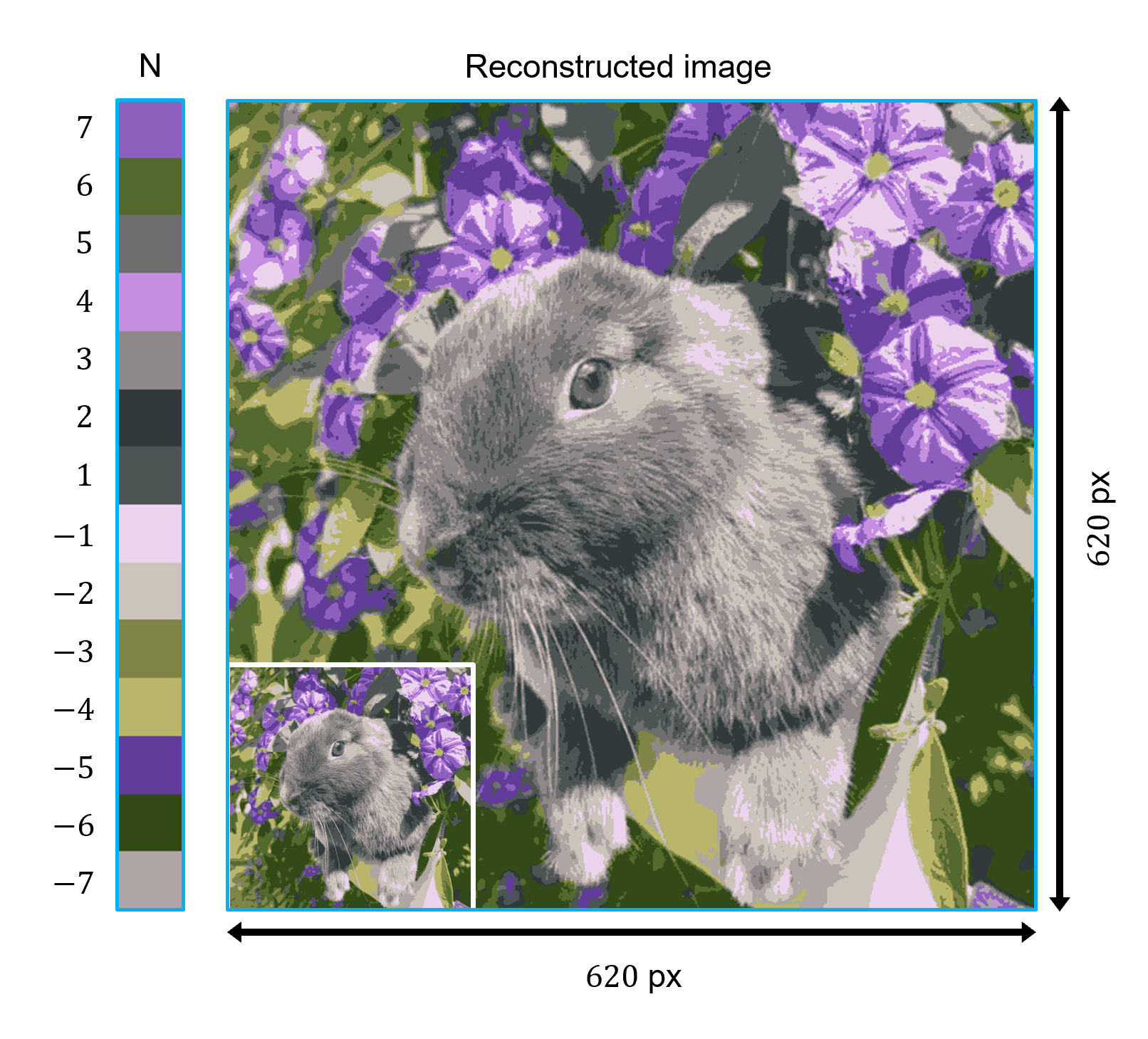}
   \caption[Image reconstruction using Skyrmion topology.]
   {\label{fig:image_reconstruction} 
    \textbf{Image reconstruction using Skyrmion topology.} The reconstructed image compiled using single-shot measurements taking by our topology detector. The image that is transmitted to the detector is shown as an inset and is constructed using a 14-levelled encoding scheme, utilising various Skyrmion topologies (N) as the information carriers.}
\end{figure}

Finally, we demonstrated the practical utility of our detector by using it to reconstruct a transmitted image encoded using a 14-levelled topological alphabet, leveraging our topological Skyrmions as information carriers.  The result of the received image is shown in Fig.~\ref{fig:image_reconstruction}, with the transmitted image shown as an inset. Each topological number was assigned a particular RGB value, creating an alphabet that was used to send the information contained within each individual pixel. This allowed us to reconstruct an image pixel-by-pixel by sending a corresponding topological state through our system to the detector, with the image being retrieved with $100\%$ accuracy.
\subsection{Noise robustness}
\begin{figure*}[ht!]
	\centering
\includegraphics[width=\linewidth]{ }
		\caption{\textbf{Topology detection through incoherent isotropic noise.} (a) Topology detection fidelity when applying increasing levels of isotropic noise. A detection accuracy of $100\%$ is maintained until the noise level reaches $-0.12$~dB, corresponding to noise strength that is $\approx 97\%$ of the average signal. (b) Varying noise floor measurements representing the ambient isotropic noise. (c) Examples of outputs captured in both channels for the approximate maximum level of $-0.12$ ~dB, showing a smeared detection probability for the values of $\ell$, with the maximum value still representing the correct output state.}
     \label{fig:isotropic_noise}
\end{figure*}
Our skyrmionic topology detector exhibits exceptional robustness to ambient noise, enabling reliable topological detection under real-world conditions where conventional Stokes polarimetry would prove challenging. We tested the robustness of our detector to incoherent fluctuations in the form of isotropic noise, a form of noise that skyrmions have proven invariant to \cite{ornelas2025topological} but can be challenging to measure using traditional approaches such as Stokes polarimetry. We see that the detector accuracy remains at 100\% from a noise level of $-17.93$~dB to $-0.12$~dB, as shown in Fig.~\ref{fig:isotropic_noise} (a). The background images taken to determine varying levels of ambient noise can be seen in Fig.~\ref{fig:isotropic_noise} (b), the dB noise level for each measurement was determined by averaging the pixel values in each image and comparing it to the average strength of the desired signals - further information on this can be found in the SI. As the noise level increased past $-0.12$~dB we observe the accuracy drop exponentially. Examples of the output fields captured for a noise level of $-0.12$~dB are shown in Fig.~\ref{fig:isotropic_noise} (c), where we can see the detection probability distributions of $\ell_{1(2)}$ spread out between all possible measurement outcomes. In our example for the state $[\ell_1,\ell_2]=[-3,1]$, we see that the average difference between the detection probability of our desired output signal compared to strongest unwanted output is $0.9\%$, however, by selecting the maximum signal from all channel measurements, this minute difference is satisfactory to accurately determine the correct input mode. This demonstrates the detectors remarkable resilience to isotropic noise, successfully allowing the measurement of the topology even when the isotropic noise peak intensity is $\approx97\%$ of the signal's, with no required noise reduction and minimal post processing. 

\section*{Discussion and conclusion}
Conventional techniques to quantify the optical topology and return a Skyrmion number include state tomography and polarimetry, both of which are highly sensitive to noise, time consuming and require extensive post-processing of the data. Our method relies on dual coincidences in two channels, each made up of an array of detectors, and returns immediately the topological number with minimal computational post-processing. This was achieved using a deep diffractive optical neural network designed to resolve both the vorticity and polarity simultaneously, enabling direct measurement of the skyrmion number. In our experiment we identify up to 15 distinct topological classes, which could be extended to a spectrum consisting of $2d-3$ distinct topological classes using $d$ detectors on each channel. While our measurement scheme is restricted to a specific family of skyrmion states, it can easily be adapted to any polarisation basis and any mode type, and points the way to more general detectors for topology.\\

Our diffractive network was synthesized using sequential free-space diffraction, interleaved with phase modulations on SLMs. We used 2 independent processing channels consisting of 5 phase layers each, with each layer decomposed into linear combinations of 121 weighted Fourier modes, thereby reducing the number of training parameters. This enabled in situ training and adaptive control of the system. For $d=9$ modes, we achieved up to $93\%$ accuracy in mode identification and a channel efficiency above $75\%$. These parameters can be further improved with increased training time, appropriate tuning of the cost function, and a modest increase in the number of phase layers. \\

We further evaluated the robustness of our approach against modal noise and spurious isotropic noise. Our results show that the detector reliably extracts the skyrmion number even at isotropic noise levels up to $-0.12$~dB. This highlights the resilience of the measurement scheme to ambient fluctuations, background light, and thermal noise, all of which typically degrade the signal-to-noise ratio. Remarkably, even under these extreme noise conditions, the detector still achieves perfect classification of the input topological states. Although our demonstration uses classical laser beams, the underlying principle is directly extendible to quantum light, enabling the detection of single-photon skyrmions \cite{ma2025nanophotonic} and non-local skyrmions \cite{ornelas2024non} with only minimal modification. A full exploration of these quantum regimes remains an exciting avenue for future work.
\section*{Methods}
\subsection*{D2NN training procedure}
Our diffractive phase screens were constructed from a weighted superposition of 2-dimensional Fourier basis modes. These modes, along with free-space propagation and a Fourier transforming lens were used to approximate our desired transmission function. The optimisation algorithm we used is known as simultaneous perturbation stochastic approximation (SPSA), which is an iterative gradient descent algorithm that utilises two sets of ansatz coefficients $\bar{\mu}_{\pm}$, along with their corresponding cost function evaluations $C(\bar{\mu}_{\pm})$, to produce a gradient vector in the direction of a desired target state, aiming to find the optimum values of $\bar{\mu}$ such that $C(\bar{\mu})$ is minimised.

To outline, the SPSA algorithm we consider the $k^{\text{th}}$ iteration during an optimisation procedure. We first produced a random step direction for each coefficient n, given by $(\Delta_k)_n \in(\pm1)_n$. Using $\Delta_k$ we could calculate the gradient as
\begin{equation}
    \hat{g}_{k}=\frac{C(\bar{\mu}_-)-C(\bar{\mu}_+)}{2\beta_{k}}\Delta_k,\label{eqn:gradient_formula} 
\end{equation}
where $\bar{\mu}_{\pm}=\bar{\mu}_k\pm\beta_k\Delta_k$ were the shifted coefficients used to sample the cost function evaluations for calculating the gradient and $\beta_k$ was a step size parameter that influenced the accuracy of the gradient calculation. This gradient was then used to estimate the coefficient ansatz for the next iteration (i.e. $k+1$) as
\begin{equation}
\mu_{k+1}=\mu_{k}+\alpha_{k}\hat{g}_{k},\label{eqn:ceofficient_update} 
\end{equation}
where the functions $\alpha_k$ is another step size parameter that determined the degree of the parameter shift in the direction of the gradient $\hat{g}_k$. The step size parameters $\alpha_k$ and $\beta_k$ influenced the convergence rate of the minimisation and therefore the accuracy of the estimated coefficients. These variables were in the form
\begin{align}
    \alpha_{k}&=\frac{a}{(k+1)^{\gamma_1}}, &\beta_k=\frac{b}{(k+1)^{\gamma_2}}\label{eqn:convergence_parameters}
\end{align}
where the values of $a$,$b$,$\gamma_1$ and $\gamma_2$ are problem dependent and are optimised through simulation. We note here that the D$^2$NN consisted of multiple phase screens, each using the same basis set but with different weighting coefficients - therefore, the number of variables needing to be solved depended on the number of phase screens being used. Regardless of this, the SPSA procedure remained unchanged and one could merely solve for all phase mask coefficients in parallel by defining a matrix of coefficients $(\mu_k)_{m,n}$ and generating a step direction matrix $(\Delta_k)_{m,n}$, where $m$ denoted the phase screen layer and $n$ denoted the individual basis mode number within that phase screen. This grouped optimisation strategy ensured that only two projective measurements were ever needed to calculate the gradient for each input state regardless of the number of variables needing to be solved. Since we optimised the mappings of multiple input states simultaneously, projective measurements were made for each input state separately within the same epoch, requiring $2d$ projective measurements per epoch, where $d$ was the number of input states being measured. The cumulative errors of all measured outputs was used to calculate the cost function, defined as
\begin{equation}
    C(\bar{\mu})=\frac{1}{d}\frac{w_1\text{MSE}+w_2\text{V}'+w_3\text{E}'}{\sum_pw_p}
\end{equation}
where the vector $\bar{w}$ weights each component in the sum to allow for the adjustment of the cost function as needed. The MSE term was defined as the sum of the mean squared errors between the output field intensity $\ket{\psi_{i}^{\text{out}}}$ and the corresponding target fields $\ket{\Phi_i}$, given by
\begin{equation}
    \text{MSE}=\sum_i^d\ |\ket{\psi_{i}^\text{out}}-\ket{\Phi_i}|^2
\end{equation}
 The visibility function $V$, quantified the detection probability distribution of the various input modes based on the inner product between the output and target intensity measurements, defined as
 \begin{equation}
     \text{V}_i=\frac{|\langle\psi_{i}^\text{out}|\phi_i\rangle|^2}{\sum_j|\langle\psi_{i}^\text{out}|\phi_j\rangle|^2},
 \end{equation}
 As such, to maximise the visibility, we minimised the complementary loss (V') of the normalised visibility
\begin{equation}
    \text{V}'=\sum_{i}^D\left(1-\text{V}_i\right).
\end{equation}
Finally, the efficiency term $E$ quantified the fraction of intensity captured within the predefined channel regions as opposed to the total intensity captured in the output image, which was obtained by calculating
\begin{equation}
    \text{E}_i = \frac{\sum|\psi_{i}^c|^2}{\sum|\ket{\psi_{i}^\text{out}}|^2}
\end{equation}
where $\sum|\psi^c_i|^2$ represented the sum of the total intensity captured within all the channels for a projection of the input field $\ket{\psi_{i}^\text{out}}$. This was again optimised by minimising the complementary loss of the efficiency (E'). given by
\begin{equation}
   \text{E}' =\sum_i^D \left(1-\text{E}_i\right).
\end{equation}
Using these projective measurements we could optimise the weighting coefficient of our basis modes, such that $C(\bar{\mu})$ was minimised.\\

Since the 2-D Fourier basis was chosen to construct our phase screens, the complex nature of the Fourier coefficients, given by $\mu_n=w_n+i\nu_n$, necessitated the need to solve both a real and imaginary coefficient for each basis component simultaneously. The real coefficient $w_n$ again served as an amplitude term and the imaginary coefficient $\nu_n$ shifted the basis modes in the frequency domain, effectively translating them in the x-y plane in a direction that depended on the orientation of the basis mode. We employed the same optimisation procedure for both coefficients in tandem by generating two separate step direction matrices and combining them to construct a complex step matrix $(\Delta_k)_{m,n}=(\Delta_{k,1})_{m,n}+i(\Delta_{k,2})_{m,n}$. Doing so allowed us to calculate a complex gradient $g_k=\hat{g}_{k,1}+i\hat{g}_{k,2}$ using the same two projective measurements to produce a new update rule in the form
\begin{equation}
    \mu_{k+1}=\mu_{k}+\alpha_{k}(\hat{g}_{k,1}+i\hat{g}_{k,2})
\end{equation}
A fail-safe procedure was implemented alongside this SPSA algorithm to account for environmental disturbances during experimental optimisation, more details on this can be found in the SI. When we experimentally applied this training procedure to the Fourier basis modes we found the optimum parameters to be $a=3\times10^{3}$, $b=3$, $\gamma_1=0.2$ and $\gamma_2=0.101$.  We used $242$ coefficients per phase screen, which corresponded to the real and imaginary terms of $121$ Fourier components arranged in a $11\times11$ grid. This grid was defined in the frequency domain and was inverse Fourier transformed to generate the final Fourier phase mask that was displayed on the SLM, with a resolution of $300 \times300$ pixels. Given our double channelled, 5 layer D$^2$NN we required a total of $2420$ trainable parameters for all 10 screens. 

\section*{Acknowledgements}
The authors would like to thank Pedro Ornelas for his valuable insight regarding the measurement of Skyrmion numbers. I.N., H.B. and A.F. would like to acknowledge the South African Quantum Technology Initiative (SA QuTI), the NRF/CSIR Rental Pool Programme and the Oppenheimer Memorial Trust for financial support.\\

\section*{Contributions}
H.B. led the implementation of the method. H.B. and C.P. and R.K. performed the experiment and analysed the data and all the authors contributed to writing the manuscript.  I.N. conceived and supervised the project.

\section*{Materials and correspondence}
Correspondence and requests for materials should be addressed to IN.

\section*{Competing financial interests}
The authors declare no financial interests.

\clearpage
\appendix

\setcounter{section}{0}
\setcounter{figure}{0}
\setcounter{table}{0}
\setcounter{equation}{0}
\setcounter{footnote}{0}
\renewcommand{\thesection}{S\arabic{section}}
\renewcommand{\thefigure}{S\arabic{figure}}
\renewcommand{\thetable}{S\arabic{table}}
\renewcommand{\theequation}{S\arabic{equation}}

\section*{Supplementary: Detailed experimental setup} 
\begin{figure*}[ht!]
	\centering
\includegraphics[width=\linewidth]{  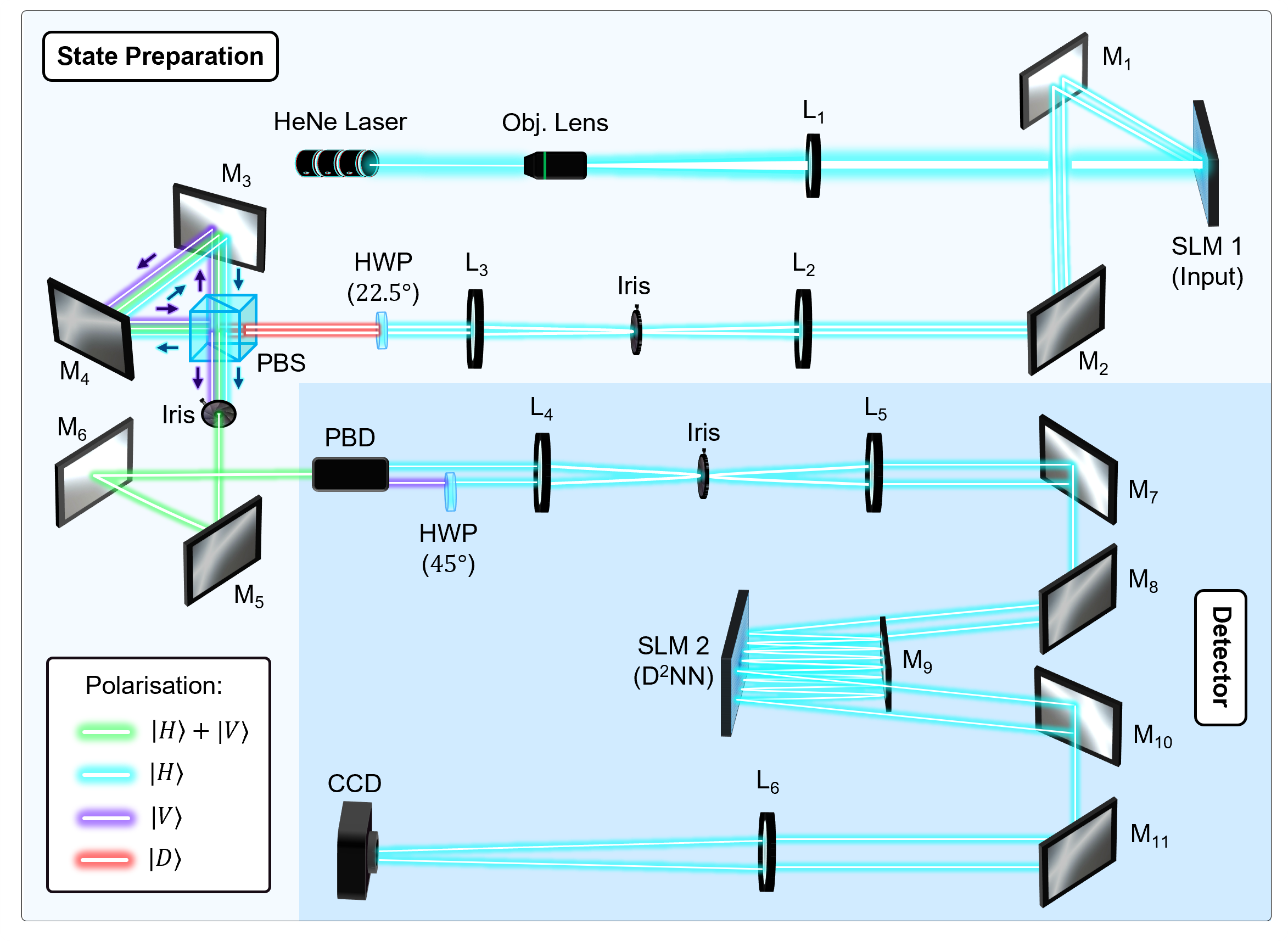}
	\caption{\textbf{Experimental set-up for deterministic Skyrmionic topology detector.} (\textbf{State Preparation}) Starting with the state preparation, a Helium-Neon laser is expanded by a 20x objective lens and subsequently collimated by a plano-convex lens L$_1$, forming a plane-wave that overfills the screen of SLM 1. On SLM 1 we encode two adjacent holograms that are imaged into a Sagnac interferometer using lenses L$_2$ and L$_3$, which an aperture placed at the focal plane of the first lens to select only the first order mode to be transmitted through the system. Before reaching the Sagnac, a half-wave plate (HWP) rotated to $22.5^{\circ}$ to convert the horizontally polarised beam to a diagonal polarisation. Within the Sagnac the two encoded modes are combined to form a Skyrmion field containing both $\ket{H}$ and $\ket{V}$ polarisations within its structured. (\textbf{Detector}) This Skyrmion field is then sent to the detection system, where it is immediately separated into its constituent components by a polarising beam displacer (PBD). Another half-wave plate, rotated to $45^{\circ}$ is placed in the way of the $\ket{V}$ polarised beam to switch its polarisation to $\ket{H}$. The input state is imaged onto the first plane of SLM 2, using lenses L$_4$ and L$_5$, which houses our Diffractive Deep Neural Network. The beam reflects back and forth between SLM 2 and a mirror M$_9$ in two separate channels, and the field at the last plane of at the SLM is Fourier transformed into a CCD camera where the intensity of each field component is captured in its allocated region.}
	\label{fig:experimental_setup}
\end{figure*}
The experimental setup is illustrated in Fig.~\ref{fig:experimental_setup} and consisted of a generation stage and detection stage. A Helium–Neon (He–Ne) laser with a wavelength of 633 nm was used to produce a coherent Gaussian beam. The beam was magnified using a $20\times$ objective lens and subsequently collimated with a plano-convex lens of focal length $f = 250$ mm, which overfilled the screen of the first spatial light modulator (SLM 1; HoloEye PLUTO-2.1-VIS, resolution 1920 × 1080 pixels). Two holograms were encoded onto each half of SLM 1 using complex amplitude modulation calculated according to \cite{arrizon2007pixelated},
\begin{equation}
    H(x,y) = J^{-1}_1 (A(x,y)) \sin[\phi(x,y) + 2\pi(G_x x + G_y y) ] \,,
\end{equation}
where $A(x,y)$ and $\phi(x,y)$ is the amplitude and phase of the desired field respectively, $J^{-1}_1$ is the inverse Bessel function of the first kind and $G_{x(y)}$ are the grating frequencies in the horizontal and vertical directions. Each hologram generated a single, horizontally polarised Laguerre–Gaussian (LG) mode. The digital nature of the holograms allowed for the topological charge of both LG modes $\ell_{1(2)}$ to be modulated independently, allowing for maximum flexibility in generating a large variety of optical skyrmions. To obtain the vectorial combination of these generated scalar modes, the resultant field from SLM 1 was imaged using a 4-f imaging system composed of two uncoated plano-convex lenses, each with a focal length of $f = 300$ mm. A circular aperture was placed at the Fourier plane between these lenses to isolate the first diffraction order of the hologram, where the desired scalar complex fields are located. After the second lens and before reaching the imaging plane, a half-wave plate (HWP) oriented at $22.5^{\circ}$ was used to convert the horizontally polarized LG modes into diagonally polarized modes. These modes were then directed into a Sagnac interferometer \cite{shen2022generation} consisting of a polarizing beam splitter (PBS) and two dielectric mirrors, as shown schematically in Fig.~\ref{fig:experimental_setup}. Inside the Sagnac interferometer, the vertically and horizontally polarised components are separated and circulated in opposite directions before being combined coaxially to generate the desired vectorial beam, with the original scalar LG modes now in the horizontal (H) and vertical (V) polarization bases. The topological numbers of the input state could be controlled by varying the encoded topological charges $\ell_1$ and $\ell_2$ on SLM 1.\\

The prepared input vectorial field was then sent to the topology detector, which consisted of a polarizing beam displacer (PBD), a second Spatial Light Modulator (SLM 2; HoloEye ERIS-1, resolution 1920 × 1200 pixels), a square dielectric mirror (size $\sim$ 15 mm), and a CCD camera was used to record the final output modes. SLM 2 and the square mirror were placed parallel to each other and separated by 50 mm, as shown in Fig.~\ref{fig:experimental_setup}, together forming a Diffractive deep neural network (D$^2$NN). Upon entering the topology detector, the beam first passed through the PBD, which separated it into two constituent scalar modes corresponding to horizontal ($\ket{H}$) and vertical ($\ket{V}$) polarizations. These orthogonal scalar modes were displaced vertically (along the y-axis) by approximately 4 mm and propagated parallel to each other along the z-axis. Since SLM 2 in the D$^2$NN was sensitive only to horizontally polarized light, a HWP oriented at $45^{\circ}$ was placed in the path of the vertically polarized beam to convert it into horizontal polarization. Now, both horizontally polarized scalar modes were imaged onto SLM 2 using a 4-f imaging system with lenses of focal length $f = 300$ mm. SLM 2 hosted the D$^2$NN, which was comprised of an upper and a lower channel, each containing 5 Fourier phase-mask layers with a resolution of $300 \times 300$ pixels per screen. The incoming modes underwent multiple reflections between SLM 2 and a square mirror (M$_9$), forming a multi-plane light converter (MPLC), as illustrated in Fig.~\ref{fig:experimental_setup}. The alignment of the MPLC was first performed manually by overlapping the beam with each phase-mask layer through fine adjustments of the mirrors M$_7$ and M$_7$. Once the coarse alignment was achieved, digital optimization was carried out by precisely shifting the phase masks to achieve a high-level of accuracy. Each phase screen was interspersed by approximately 100~mm of free-space propagation between SLM 2 and the square mirror, a distance that remained within the Rayleigh range of the beam. To suppress unmodulated beam intensity, a linear grating was applied along the x-direction in combination with the Fourier phase masks up to the fourth layer. In the fifth (final) layer, gratings were applied in both x and y directions to extend the spatial profile of the beam, thereby minimizing overlap and interference in the far field.

The output field that emerged from the final phase screen of the D$^2$NN was Fourier-transformed by a 500~mm lens, and the resulting far-field intensity distribution was recorded using a CCD camera. The CCD plane was divided into two output regions corresponding to Channel 1 (the $\ket{H}$ component) and Channel 2 (the $\ket{V}$ component). Within each region, the target modes were defined by spatially separated Gaussian spots. By tracing these regions, output channels corresponding to each input state were established. The intensity profile corresponding to each polarization component was captured in their respective region, and the value of $\ell$ for each component was determined through a normalized inner product between the output field and all target states. The output channel with the maximum inner product was taken as the detected $\ell$ for that polarization. By determining the values of $\ell_1$ and $\ell_2$ for the horizontal ($\ket{H}$) and vertical ($\ket{V}$) components, respectively, we could reconstruct the full input topological field.

\section*{Supplementary: Measuring the skyrmion number with Stokes polarimetry} 
We obtain the skyrmion number, $N$, from experimental measurements to independently confirm the accuracy of our detector, we make use of the approach proposed by McWilliam \textit{et al.} in Ref.\cite{mcwilliam2023topological}. This approach makes use of a contour integral over complex polarisation fields to compute the wrapping number,
\begin{equation}
N = \frac{1}{4 \pi} \int_{A} \mathbf{S} \cdot 
\left( 
\frac{\partial \mathbf{S}}{\partial x} \times 
\frac{\partial \mathbf{S}}{\partial y}
\right)\, dx\, dy
\label{eq:SkyrmeNo}
\end{equation}
We build our measurement from these result,
\begin{equation} \label{eq:line_int}
     N = \frac{1}{2} \left( \sum_{j} S_{z}^{(j)} N_{j} - \bar{S_{Z}^{\infty}} N_{\infty} \right) \,,
\end{equation}
where $N_j$ is the charge of individual phase singularity at position $j$ in the field $S_x + iS_y $,  $S_z^{(j)}$ is the value of the Stokes parameter $S_z$ at the point $j$, $N_{\infty}$ is the result of the contour integral at infinity and $S_z^{(\infty)}$ is the value of the Stokes parameter $S_z$ as $r\rightarrow \infty$. The Stokes parameters were experimentally determined by measuring six polarisation intensity projections,
\begin{equation}
\mathbf{S} =
\begin{pmatrix}
S_0 \\[2pt]
S_1 \\[2pt]
S_2 \\[2pt]
S_3
\end{pmatrix}
=
\begin{pmatrix}
|E_x|^2 + |E_y|^2 \\[4pt]
|E_x|^2 - |E_y|^2 \\[4pt]
2\,\mathrm{Re}(E_x E_y^{*}) \\[4pt]
2\,\mathrm{Im}(E_x E_y^{*})
\end{pmatrix}
=
\begin{pmatrix}
I_H + I_V \\[4pt]
I_H - I_V \\[4pt]
I_D - I_A \\[4pt]
I_R - I_L
\end{pmatrix}.
\label{eq:stokes}
\end{equation}
where the subscripts H, V, D, A, R and L represent horizontal, vertical, diagonal, anti-diagonal, right circular and left circular polarisations respectively. A linear polariser and half-wave plate was used to acquire the intensity projections for the linear projections and a linear polariser in combination with a quarter-wave plate  and half-wave plate for the circular projections \cite{singh2020digital}. Eq. \ref{eq:SkyrmeNo} requires the locally normalised Stokes parameters $S_j$ which were computed from the experimentally obtained Stokes parameters $s_j$ according to,
\begin{equation}
    S_j =\frac{s_j}{\sqrt{s_1^2+s_2^2+s_3^2}} \,.
\end{equation}
Eq. \ref{eq:line_int} requires the location and charge of the singularities $N_j$ of the polarisation fields $P(x,y)$ given by,
\begin{eqnarray} \label{eq:polphase1}
P(x,y) &=& S_{2}(x,y) + iS_{3}(x,y)
\end{eqnarray}
Once this field was calculated, the locations and charges of the individual singularities were computed using a numerical equivalent to the curl $\nabla \times$ operation proposed in Ref. \cite{chen2007detection} termed the circulation $D$. The circulation is defined as,
\begin{eqnarray}
    D^{m,n} = &\frac{d}{2}& ( G_x^{m,n} + G_x^{m,n+1} +G_y^{m,n+1} + G_y^{m+1,n+1} \nonumber \\ &-  &G_x^{m+1,n+1} - G_x^{m+1,n} -G_y^{m+1,n} \nonumber \\
    &-& G_y^{m,n})\,.
\end{eqnarray} 
Here, $D^{m,n}$ represents the value of the circulation of the pixel in the $n$-th row and $m$-th column. $G_x^{m,n}$ and $G_y^{m,n}$ are the phase gradient in the horizontal and vertical direction of the pixel in the $n$-th row and $m$-th column, respectively and $d$ is the pixel size. Typically, the circulation will return a 0 value if there is no singularity at that pixel and a nonzero value if there is. The magnitude of the circulation indicates the charge of the singularity and the sign indicates the direction/handedness of the singularity. These values were then substituted into Eq. \ref{eq:line_int} to calculate the skyrmion number. The value of $S_z$ at infinity $S_z^{(\infty)}$. was computed by calculating the average value of the experimentally measured parameter around the periphery of the array. The value $N_{\infty}$ were is computed by performing a numerical contour integral around the periphery of the field.

\section*{Supplementary: characterisation of input states} \label{sec:experimental_determination}
The skyrmion beam emerging from the Sagnac interferometer exhibits a spatially varying polarization structure, which we quantified through Stokes polarimetry \cite{shen2022generation}. To obtain a complete description of the local polarization at each point across the transverse plane, we recorded six polarization-resolved intensity projections: $I_{H}$, $I_{V}$, $I_{D}$, $I_{A}$, $I_{R}$, and $I_{L}$, corresponding to horizontal, vertical, diagonal, anti-diagonal, right-circular, and left-circular polarization states, respectively \cite{gupta2015wave,hakobyan2025q}. These measurements allow the Stokes parameters to be evaluated using Eq.~\ref{eq:stokes}.
\begin{figure*}[ht!]
	\centering
\includegraphics[width=\linewidth]{ 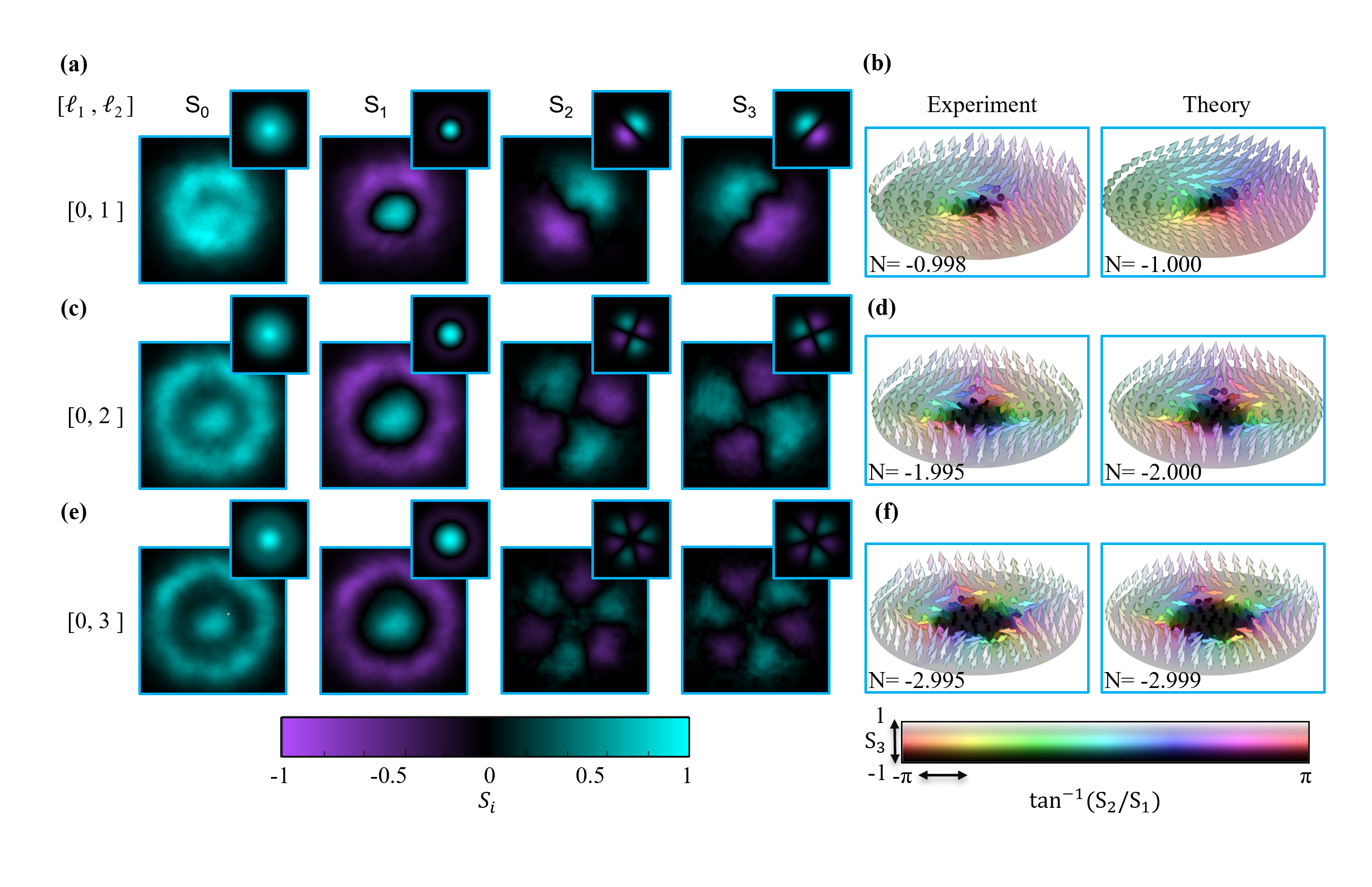}
	\caption{\textbf{Reconstruction of skyrmionic textures with negative topological charge from Stokes polarimetry.} The spatial polarization structure of vector beams is experimentally and theoretically reconstructed from full Stokes parameter measurements ($S_{0}$--$S_{3}$), obtained using six polarization-resolved intensity projections ($I_{H}$, $I_{V}$, $I_{D}$, $I_{A}$, $I_{R}$, and $I_{L}$,). For each input state, the measured Stokes distributions (left) are compared with their theoretical predictions (insets), and the corresponding polarization vector fields are shown for experiment and theory. (a-b) Input state $[l_1:l_2] = [0:1]$, yielding a skyrmion number of $N=-0.998$ experimentally and $N=-1.000$ theoretically. (c-d) Input state $[0:2]$, giving $N=-1.995$ (experiment) and $N=-2.000$ (theory). (e-f) Input state $[0:3]$, giving $N=-2.995$ (experiment) and $N=-2.999$ (theory). The topological charge is computed numerically using the line-integral method, both experimentally and theoretically \cite{mcwilliam2023topological}}
     \label{fig:Stokes_meas_negative}
\end{figure*}
Before directing the beam into the topological detection stage, we prepared $81$ topological states of the general form $\ket{\psi_{\ell_{12}}} = \ket{\ell_1}\ket{H} + e^{i\gamma}\ket{\ell_2}\ket{V}$, where $\ell_1$ and $\ell_2$ denote the orbital angular momentum (OAM) values carried by the horizontal and vertical polarization components, respectively, and $\gamma$ is the relative phase \cite{shen2024optical}. By using all possible OAM-mode combinations in the range $\ell_{1,2}\in[-4,4]$ for the $\ket{H}$ and $\ket{V}$ bases, the corresponding topological numbers lie within $N\in[-7,7]$. Among these $81$ topological states, we present three negative and three positive states as representative examples.
\begin{figure*}[ht!]
	\centering
\includegraphics[width=\linewidth]{  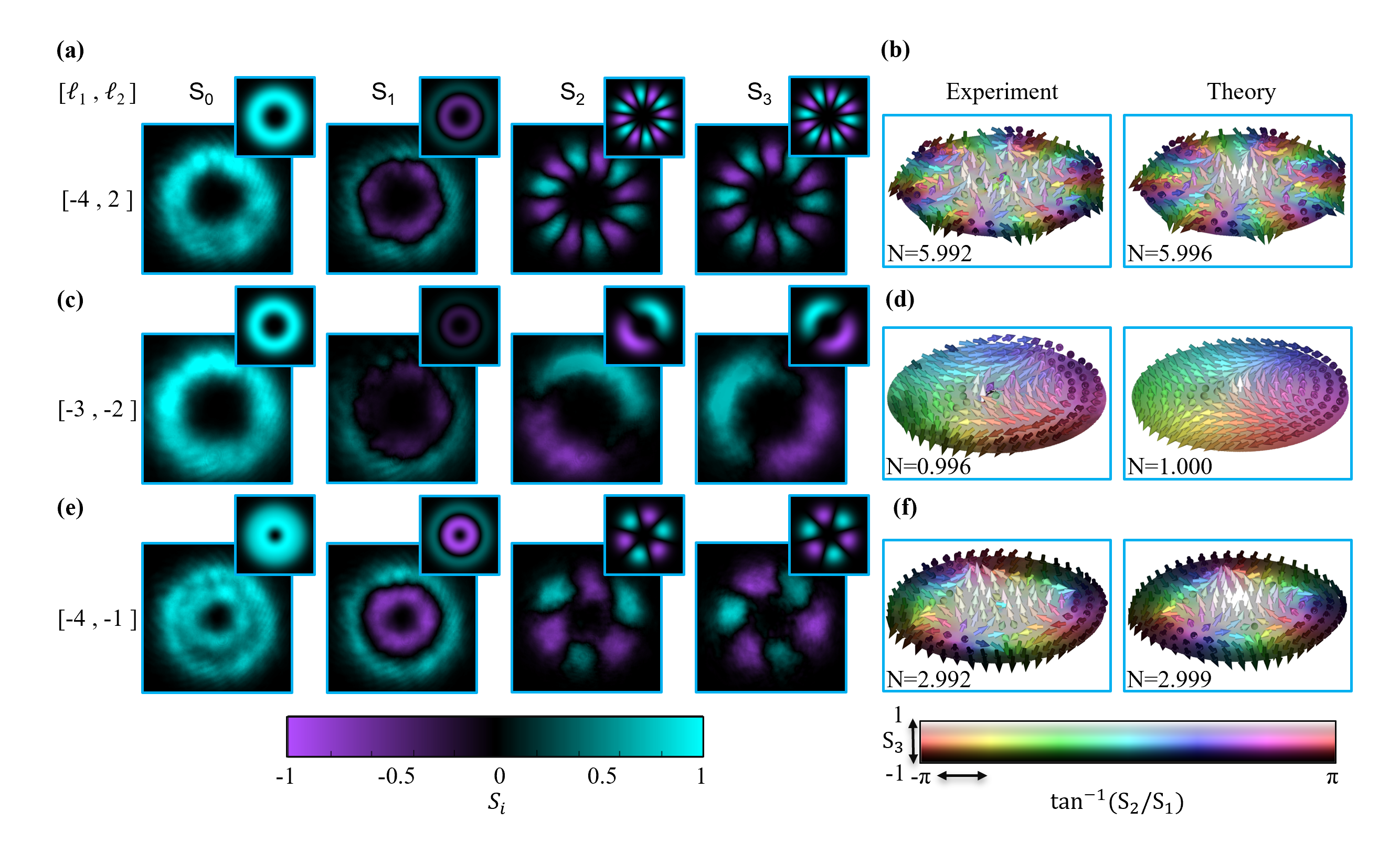}
		\caption{\textbf{Reconstruction of skyrmionic textures with positive topological charge from Stokes polarimetry.} The spatial polarization structure of vector beams is experimentally and theoretically reconstructed from full Stokes parameter measurements ($S_{0}$--$S_{3}$), obtained using six polarization-resolved intensity projections ($I_{H}$, $I_{V}$, $I_{D}$, $I_{A}$, $I_{R}$, and $I_{L}$,). For each input state, the measured Stokes distributions (left) are compared with their theoretical predictions (insets), and the corresponding polarization vector fields are shown for experiment and theory. (a-b) Input state $[l_1:l_2] = [-4:2]$, yielding a skyrmion number of $N=5.992$ experimentally and $N=5.996$ theoretically. (c-d) Input state $[-3:-2]$, giving $N=0.996$ (experiment) and $N=1.000$ (theory). (e-f) Input state $[-4:-1]$, giving $N=2.992$ (experiment) and $N=2.999$ (theory).}
     \label{fig:Stokes_meas_positive}
\end{figure*}\\

Here, we computed the topological charge using the line-integral method detailed above, both experimentally and theoretically. In Fig.~\ref{fig:Stokes_meas_negative}(a–f), we show the negative topological states. Following Eq.~\ref{eq:stokes}, for each row, the first panel [(a), (c), (e)] presents the experimental Stokes vector distribution compared with their theoretical predictions (insets), while the second panel [(b), (d), (f)] shows the corresponding skyrmionic texture of the fields. Figures (a–b) correspond to the input state $\ket{0}\ket{H} + e^{i\pi/4}\ket{1}\ket{V}$, for which the experimentally measured and theoretical topological numbers are $-0.998$ and $-1.000$, respectively.  
Figures (c–d) correspond to the input state $\ket{0}\ket{H} + e^{i5\pi/4}\ket{2}\ket{V}$, with experimental and theoretical values of $-1.995$ and $-2.000$, respectively.  
Figures (e–f) correspond to the input state $\ket{0}\ket{H} + e^{i\pi/4}\ket{3}\ket{V}$, with experimental and theoretical values of $-2.995$ and $-2.999$, respectively.\\

Similarly, in Fig.~\ref{fig:Stokes_meas_positive}(a–f), we present the positive topological states. As before, the first panel in each row [(a), (c), (e)] shows the experimental Stokes vector distribution, whereas the second panel [(b), (d), (f)] illustrates the corresponding skyrmionic textures of the fields. Figures (a–b) correspond to the input state $\ket{-4}\ket{H} + e^{i5\pi/4}\ket{2}\ket{V}$, with experimentally measured and theoretical topological numbers of $5.992$ and $5.996$, respectively. Figures (c–d) show the input state $\ket{-3}\ket{H} + e^{i\pi/4}\ket{-2}\ket{V}$, having experimental and theoretical values of $0.996$ and $1.000$, respectively. Figures (e–f) correspond to the state $\ket{-4}\ket{H} + e^{i\pi/4}\ket{-1}\ket{V}$, for which the experimental and theoretical topological charges are $2.992$ and $2.999$, respectively.\\

The RGB colour map of the Stokes vectors in the skyrmionic texture plots of Figs.~\ref{fig:Stokes_meas_negative} and \ref{fig:Stokes_meas_positive} illustrates the transverse distribution of the vector orientations, given by $\tan^{-1}(S_2/S_1)$, as well as the polarity encoded in $S_3$ \cite{shen2022generation,shen2024optical}. The polarity of each topological state is indicated by the orientation of the polarization vectors at the centre and boundary of the texture, with values ranging between $+1$ and $-1$. In the plots, white-coloured vectors pointing upward indicate a polarity of $+1$, whereas black-coloured vectors pointing downward depict a polarity of $-1$. In Figs.~\ref{fig:Stokes_meas_negative}(b), (d), and (f), the vectors at the beam centre predominantly point downward, indicating a polarity of $-1$, while the boundary vectors mostly point upward, corresponding to $+1$ polarity. In contrast, in Figs.~\ref{fig:Stokes_meas_positive}(b), (d), and (f), the centre vectors largely point upward, indicating a polarity of $+1$, while the boundary vectors predominantly point downward, corresponding to $-1$ polarity~\cite{shen2022generation}.\newpage

\section*{Supplementary: Further details on noise robustness}
\label{sec:noise_robustness}
We subjected our system to incoherent noise with varying degrees of strength to test its resilience. The incoherent noise was added to the system in the form of an isotropic tunable white light source aimed directly into the detectors CCD camera at the output plane. We measured the average detection accuracy for a full set of 81 topological states, constituting every possible state combination for our trained topological charge range of $\ell=[-4,4]$, for increasing noise levels, The noise levels ranged from $-17.93$~dB to $-0.016$~dB, where the noise floor was calculated by averaging all pixel values from a camera image taken with no mode projections, representing the ambient environmental noise. The noise level was found by taking the ratio of the noise floor to the strength of the signal found in the desired channels, averaged over all 81 measurements and both channels. The topology detection accuracy remained at $100\%$ up to $\approx -0.12$~dB, which corresponded to a noise level with a strength of $\approx97\%$ of that of the average signal, i.e.
\begin{equation}
    \text{dB} =10\times \log_{10}\left (\frac{\text{noise}}{\text{signal}}\right).
\end{equation}
The probability distribution used to detect the corresponding values of $\ell_1$ and $\ell_2$ are found using our visibility measure
\begin{equation}
    \text{V}_i=\frac{|\langle\psi_{i}^\text{out}|\phi_i\rangle|^2}{\sum_j|\langle\psi_{i}^\text{out}|\phi_j\rangle|^2}, \label{eqn:visibility}
\end{equation}
where $\ket{\psi^{\text{out}}_i}$ are our experimentally obtained output states and $\ket{\phi_j}$ are the various simulated target states for each channel. This operation resembles the normalised coupling strength of our output signal into a bundle of single mode fibres. We take the maximum coupled output mode as our detected state, a method that is commonly used when performing optical field classification tasks \cite{li2021all,Ozcan_2018_all_optical_ML}, allowing us to correctly classify our incoming light fields $\ket{\psi^{\text{in}}_i}$, regardless of high-levels of ambient noise and with requiring no additional noise reduction and minimal post-processing. This illustrates the advantage of our approach over standard Stokes polarimetric methods, which would usually require complex post-processing procedures or multiple measurements to overcome these effects.

\section*{Supplementary: Fail-safe procedure}
\label{sec:failsafe_procedure}
We implemented a "fail-safe" procedure alongside the SPSA optimisation to account for environmental perturbations to the system during experimental training. The fail-safe routine operated by periodically tracking the minimum cost value encountered during optimisation, together with its associated phase screen coefficient vector. Every $t=5$ iterations, the cost evaluation in the current step $C(\bar{\mu}_\text{current})$ was compared against this stored minimum $C(\bar{\mu}_\text{best})$. We introduced a convergence criterion $\epsilon=5\times10^{-4}$, such that if the improvement in the cost function over the interval $t$ fell below this threshold, i.e. $C(\bar{\mu}_\text{best})-C(\bar{\mu}_\text{current})<\epsilon$, the optimisation landscape was deliberately perturbed by modifying the cost function weightings $\bar{w}$, according to 
\begin{equation}
    \bar{w}=\frac{\bar{w}_i}{[1,\hspace{0.2cm}V,\hspace{0.2cm}E]}+[1,\hspace{0.2cm}0,\hspace{0.2cm}0]\cdot(-1)^{\epsilon'}, \label{eqn:cost_function_change}
\end{equation}
where V and E denote the visibility and efficiency metrics, $\bar{w}_i=[1,3,1]$ are the initial cost function weightings and $\epsilon'$ counts the number of times the stagnation criterion was triggered. The alternating term being added in Eq.~\ref{eqn:cost_function_change} effectively toggles the contribution of the mean squared error component, cyclically tightening and relaxing the emphasis placed on achieving the desired Gaussian intensity profile. This adaptive reshaping of the optimisation landscape disrupts the stagnation, helping the algorithm escape shallow plateaus and explore previously inaccessible regions of the parameter space. Its impact is evident in the oscillatory structure of the cost function curve during training, where the characteristic peaks and troughs mark each intervention of the fail-safe mechanism. Additionally, during the fail-safe check, the ambient noise floor is measured by capturing a blank image with the CCD (i.e. an image measurement without any mode being projected). The average pixel value from this image is calculated and deducted from any projective measurement taken during the training process. This allows for the system to adapt to background noise fluctuations, keeping the training behaviour of the detection visibility consistent amidst imperfect laboratory conditions.
\end{document}